\documentclass[aps,prd,preprint,nofootinbib,11pt]{revtex4}
\usepackage{amsfonts}
\usepackage{mathrsfs}
\usepackage{graphicx}
\usepackage{amsmath}
\usepackage{amssymb}
\usepackage{multirow}
\usepackage{subfigure}
\usepackage{epsfig}
\usepackage{graphicx}
\usepackage{color}
\newcommand{\bqa}{\begin{eqnarray}}
\newcommand{\eqa}{\end{eqnarray}}
\newcommand{\beq}{\begin{equation}}
\newcommand{\eeq}{\end{equation}}
\allowdisplaybreaks[1]
\graphicspath{{fig/}{dia/}} \DeclareGraphicsExtensions{.eps}
\begin{document}

\title{Mass predictions of triply heavy hybrid baryons via QCD sum rules\\[0.7cm]}
\author{Yi-Cheng Zhao$^{1}$, Chun-Meng Tang$^{1}$, and Liang Tang$^{1, 2}$\footnote{tangl@hebtu.edu.cn}}

\affiliation{$^1$ College of Physics, Hebei Normal University, Shijiazhuang 050024, China\\
$^2$ Hebei Key Laboratory of Photophysics Research and Application, Hebei Normal University, Shijiazhuang 050024, China
}



\begin{abstract}
\vspace{0.3cm}
In this article, we study the mass spectrum of the low-lying triply heavy hybrid baryon, which consists of three valence heavy quarks in a color octet and one valence gluon, with spin-parity $J^P=(\frac{1}{2})^+$ via QCD sum rules. This is the first study on the triply heavy hybrid baryons in the framework of QCD sum rules. After performing the QCD sum rule analysis, we find that the mass of $cccg$ hybrid baryon lies in $M_{cccg}= 5.91-6.13$ GeV. As a byproduct, the mass of the triply bottom hybrid baryon state is extracted to be around $M_{bbbg}=14.62-14.82$ GeV. The contributions up to dimension eight at the leading order of $\alpha_s$ (LO) in the operator product expansion are taken into account in the calculation. The triply charmed hybrid baryon predicted in this work can decay into one doubly charmed baryon and one charmed meson. Especially, we propose to search for $cccg$ hybrid baryon with $J^{P}= (1/2)^+$ in the P-wave decay channels $\Xi_{cc}^{++}  D^0$, $\Xi_{cc}^{+}  D^+$, and $\Xi_{ccs}^{+}  D_s^+$, which may be accessible in future BelleII, Super-B, PANDA, and LHCb experiments.
\end{abstract}
 \maketitle
\newpage

\section{Introduction}

The quark model has been proven to be very successful in the classification of hadrons, the calculation of hadron spectra, and their other properties. However, hadron states which can not be accommodated in the conventional quark model were also proposed by Gell-Mann~\cite{Gell-Mann:1964ewy} and Zweig~\cite{Zweig:1964ruk} in 1964, which are named as exotic states including multiquark states, glueballs, and hybrid hadrons. On the other hand, Quantum Chromodynamics (QCD), which is widely accepted as the fundamental theory of strong interaction, allows the existence of those exotic states. Exploring the existence and properties of such exotic states is one of the most intriguing research topics of hadron physics.

Hybrid hadrons can be divided into hybrid mesons consisting of one quark, one anti-quark, and one valence gluon ($q \bar{q} g$), and hybrid baryons composed of three quarks and one gluon ($qqqg$). These are typical exotic states which contain gluonic excitations. The existence of gluonic excitation in the hadron spectrum is one of the most important unanswered questions in hadron physics. Since some exotic hybrid mesons that have quantum numbers $J^{PC}$ not available within a regular $q\bar{q}$ valence structure can be identified by their quantum numbers, a great deal of work has been done on hybrid mesons. But for baryons, the regular $qqq$ can have all the $J^P$ values. So hybrid baryons will always have to ``share" quantum numbers with regular states, thus making them very difficult to identify experimentally, although there are clear theoretical definitions. Therefore, one must use other features of these hybrid baryons to determine whether or not they are existent (for more information see Refs.~\cite{Barnes:2000vn, Page:2000fq, Page:2002mt, Lanza:2017qel}). Despite decades of searching, both hybrid mesons and hybrid baryons have not been conclusively identified in experiments yet.

For many decades there has been a great interest in detecting and studying hybrid baryons. For example, they have been studied by the bage model~\cite{Barnes:1982fj, Barnes:1983xx, Golowich:1982kx, Duck:1983ju, Carlson:1983xr}, the flux tube model~\cite{Capstick:1999qq, Capstick:2002wm}, models of the $P_{11}(1440)$ (Roper) resonance as a hybrid~\cite{Li:1991sh, Li:1991yba},  the large $N_c$ QCD~\cite{Chow:1998tq}, and the QCD sum rules~\cite{Martynenko:1991pc, Kisslinger:1995yw, Kisslinger:1998xa, Kisslinger:2003hk, Kisslinger:2009dr, Azizi:2017xyx}. It should be noted that all these researches are concentrated on the light quark hybrid baryons.

In recent years, a large number of new hadronic states containing heavy quarks (the charm quark $c$ or bottom quark $b$) have been observed at hadron colliders and $e^+ e^-$ colliders~\cite{ParticleDataGroup:2022pth}. For instance, LHCb collaboration observed a highly significant structure in the $\Lambda_c^+ K^- \pi^+ \pi^+$ mass spectrum, which is interpreted as the doubly charmed baryon $\Xi_{cc}^{++}$~\cite{LHCb:2017iph} with mass $3621 \pm 0.72 \pm 0.27 \pm 0.14$ MeV. Heavy baryon spectroscopy is experiencing rapid advancement, and it seems rather promising to establish triply heavy baryons~\cite{GomshiNobary:2003sf, GomshiNobary:2004mq, GomshiNobary:2005ur, GomshiNobary:2007xk, GomshiNobary:2007ofo} and triply heavy hybrid baryons in the near future. In the literature, the triply heavy baryons have been studied in QCD sum rules~\cite{Zhang:2009re, Wang:2011ae, Wang:2020avt, Aliev:2012tt, Aliev:2014lxa}. As mentioned in the last paragraph, we notice that no one has studied the triply heavy hybrid baryons yet. So, the purpose of this paper is to extend the study of hybrid baryons to the triply heavy sector in the framework of QCD sum rules.

Although the QCD theory has been commonly accepted as the correct theory underlying strong interaction, many questions concerning the dynamics of the quark and gluons at a large distance remain unanswered due to the effects of the low-energy QCD. The triply heavy hybrid baryon, wherever light quarks are absent, is of great interest in understanding the dynamics of QCD at the hadronic scale. They all belong to the exotic states, which cannot be explained in the traditional quark model, and they are of particular importance to understanding the low-energy behaviors of QCD, where the nonperturbative effect plays an important role in describing the properties of the hadrons.

Among those theoretical methods in dealing with the nonperturbative effects, QCD sum rules is a remarkably successful and powerful technique for the computation of hadronic properties~\cite{Shifman:1978bx, Shifman:1978by, Reinders:1984sr, Narison:1989aq, P.Col}. It is a QCD-based theoretical framework that
incorporates nonperturbative effects universally order by order using the operator product expansion (OPE). In this approach, to establish the sum rules, the first step is to construct the proper interpolating current corresponding to the hadron of interest, which possesses the foremost information about the concerned hadron, such as the quantum numbers, the constituent quarks, and gluons. By using the current, one can then construct the two-point correlation function, which can be investigated at both quark-gluon and hadron levels, usually called the QCD and the phenomenological representations, respectively. After performing the Borel transformation on both representations, we can formally establish the QCD sum rules, from which we can extract the mass of the concerned hadron.

The rest of the paper is arranged as follows. After the introduction, some primary formulas of the QCD sum rules in our calculation are presented in Sec. \ref{Formalism}. The numerical analyses and results are given in Sec. \ref{Numerical}. Section \ref{decay} is devoted to the decay analyses of the triply charmed hybrid baryon. The last part is left for conclusions and discussion of the results.

\section{Formalism}\label{Formalism}

The starting point of the QCD sum rule method is to construct the interpolating current properly and then write down the correlation function. The current of the hybrid baryon state is built up with three quarks and one gluon. In principle, the current operator for $J^P=\frac{1}{2}^{+}$ hybrid state is not unique. Therefore, various forms of light hybrid baryons have been discussed in Refs.~\cite{Martynenko:1991pc, Braun:1992jp}. In this work, as a pioneer work, we intend to study the lowest-lying triply heavy hybrid baryon with spin-parity $J^P=\frac{1}{2}^{+}$ (nominated as $\Omega_H$), so we use the form
\begin{eqnarray}
j_{\Omega_{H}}(x) &=& g_{s} \varepsilon ^{abc}[Q_{a}^{T}(x)C \gamma^{\mu} Q_{b}(x)] \gamma^{\alpha} G_{\mu\alpha}^{A}(x)[t^{A}_{cd} Q_{d}(x)] , \label{current}
\end{eqnarray}
whose structure has been extensively used in the studies of the lowest-lying light hybrid states~\cite{Kisslinger:1995yw, Kisslinger:2003hk, Azizi:2017xyx}. Here $g_{s}$ is the strong coupling constant, the subscripts $a, b, c, d= 1, 2, 3$ denote the color indices, the index $T$ means matrix transposition, $C$ is the charge conjugation matrix, $t^{A}=\lambda^{A}/2$ where  $\lambda^{A}$ is the Gell-Mann matrix, and $Q$ represents the heavy quark $c$ or $b$. The corresponding correlator is
\begin{eqnarray}
\Pi(q) &=& i \int d^4 x e^{i q \cdot x}\langle 0 | T \{j_{\Omega_H}(x), \bar{j}_{\Omega_H}(0)\} |0 \rangle .
\end{eqnarray}
According to the Lorentz covariance, the correlator has the form
\begin{eqnarray}
\Pi(q)=q\!\!\!\slash \Pi_1(q^2) + \Pi_2(q^2). \label{Lorentz-structure}
\end{eqnarray}
For each invariant function $\Pi_{1}(q^2)$ and $\Pi_{2}(q^2)$, a sum rule can be obtained.

The correlation function $\Pi_i(q^2)$ ($i=1,2$) can be investigated at both quark-gluon and hadron levels, usually called the QCD and the phenomenological representations, respectively. At the quark-gluon level, the correlation function can be calculated with the operator product expansion (OPE). In our evaluation, the heavy-quark ($Q=c$ or $b$) propagator $S^Q_{j k}(p)$ is considered in momentum space, which can be expanded as
\begin{eqnarray}
S^Q_{j k}(p) & = &\frac{i \delta_{j k}(p\!\!\!\slash + m_Q)}{p^2 - m_Q^2} - \frac{i}{4} \frac{t^a_{j k} G^a_{\alpha\beta} }{(p^2 - m_Q^2)^2} [\sigma^{\alpha \beta}
(p\!\!\!\slash + m_Q)
+ (p\!\!\!\slash + m_Q) \sigma^{\alpha \beta}] \nonumber \\ &+& \frac{i\delta_{jk}m_Q  \langle g_s^2 G^2\rangle}{12(p^2 - m_Q^2)^3}\bigg[ 1 + \frac{m_Q (p\!\!\!\slash + m_Q)}{p^2 - m_Q^2} \bigg] \nonumber \\ &+& \frac{i \delta_{j k}}{48} \bigg\{ \frac{(p\!\!\!\slash +
m_Q) [p\!\!\!\slash (p^2 - 3 m_Q^2) + 2 m_Q (2 p^2 - m_Q^2)] }{(p^2 - m_Q^2)^6}
\times (p\!\!\!\slash + m_Q)\bigg\} \langle g_s^3 G^3 \rangle \; .
\end{eqnarray}

Here, the subscripts $j$ and $k$ represent the color indices of the heavy quarks, and the vacuum condensates are displayed clearly. We refer to Refs.~\cite{Wang:2013vex, Albuquerque:2012jbz} for a detailed discussion of the heavy quark propagator.

Moreover, the perturbative propagator between the gluon field strength tensors $G_{\mu \nu}^a(x)$ and $G_{\rho \sigma}^b(0)$ employed in our analytical calculation is considered in coordinate space, which can be expressed as~\cite{Govaerts:1984hc}:
\begin{eqnarray}
 S_{\mu\nu,\rho\sigma}^{ab}(x)&=& \frac{\delta^{ab}}{2\pi^{2}} \times\frac{1}{x^{6}}\big\{(g_{\mu\rho}x^{2} - 4x_{\mu}x_{\rho})g_{\nu\sigma} - (g_{\mu\sigma}x^{2} - 4x_{\mu}x_{\sigma})g_{\rho\nu}\nonumber\\
 &-& (g_{\rho\nu}x^{2} - 4x_{\rho}x_{\nu})g_{\mu\sigma} + (g_{\nu\sigma}x^{2} - 4x_{\nu}x_{\sigma})g_{\rho\mu}\big\}.\label{pert-gluon}
\end{eqnarray}

For extracting reliable results from the comparison between the two representations of the correlation function, one should guarantee a good OPE convergence on the QCD side and simultaneously suppress the contributions from higher excited states and the continuum states on the phenomenological side. A practical way of doing this is to utilize the Borel transformation, whose definition is given by:
\begin{equation}	
  {\cal B}\! \left[ \Pi(Q^2) \right] \equiv \Pi(M_B^2) =
  \lim\limits_{\tiny \begin{matrix} Q^2, n \rightarrow \infty \\ Q^2/n = M_B^2 \end{matrix}}
  \frac{(-1)^n (Q^2)^{n+1}}{n!} \left( \frac{\partial}{\partial Q^2} \right)^n \!\Pi(Q^2)
\end{equation}
where $Q^2$ is the four-momentum of the hadron in the Euclidean space $(Q^2 = -q^2)$,
and $M_B^2$ is a free parameter of the sum rules.

In the OPE side, the correlation function $\Pi^{\text{OPE}}(q)$ can be decomposed over the Lorentz structure $\sim q\!\!\!\slash$ and $\sim I$ as shown in Eq.(\ref{Lorentz-structure}). In calculations, we choose the term $\sim q\!\!\!\slash$~\cite{Wang:2017qvg, Azizi:2017xyx}. Then the correlation function $\Pi_1(q^2)$ can be expressed in terms of a dispersion relation as
\begin{eqnarray}
  \Pi^{\text{OPE}}_1 (q^2) = \int_{(3 m_Q)^2}^\infty ds \frac{\rho_{1}^{\text{OPE}}(s)}{s - q^2}, \label{Pi-OPE}
\end{eqnarray}
where the spectral density $\rho_1^{\text{OPE}}(s)$ has the following relation: $\rho_{1}^{\text{OPE}}(s) = \text{Im} [\Pi_{1}^{\text{OPE}}(s + i \varepsilon)]/\pi$. It can be expanded as follows
\begin{eqnarray}
  \rho_{1}^{\text{OPE}}(s) &=& \rho_{1}^{\text{pert}}(s) + \rho_{1}^{\langle G^2 \rangle}(s) + \rho_{1}^{\langle G^3 \rangle}(s) + \rho_{1}^{\langle G^4 \rangle}(s)\, , \label{spectral-density}
\end{eqnarray}
where $\rho_{1}^{\text{pert}}(s)$, $\rho_1^{G^2}(s)$, $\rho_1^{G^3}(s)$, and $\rho_1^{G^4}(s)$ represent the spectral densities of the perturbative term, the two-gluon condensate contribution, the three-gluon condensate contribution, and the four-gluon condensate contribution, respectively.

\begin{figure}[htb]
\begin{center}
\includegraphics[width=10cm]{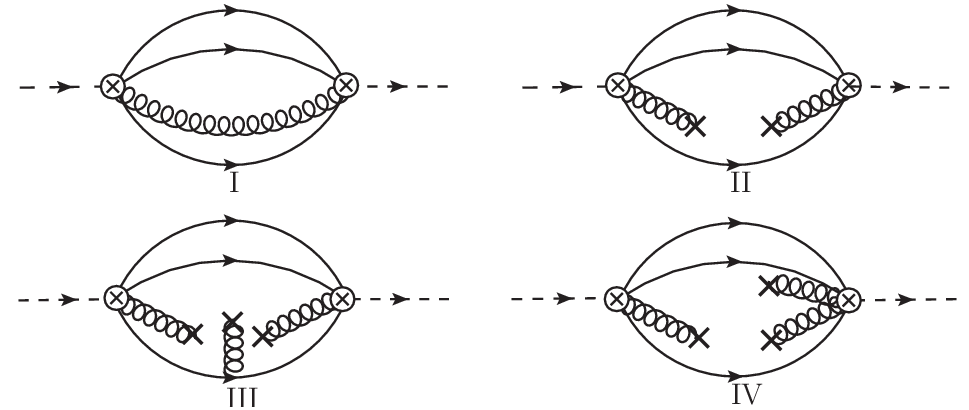}
\includegraphics[width=10cm]{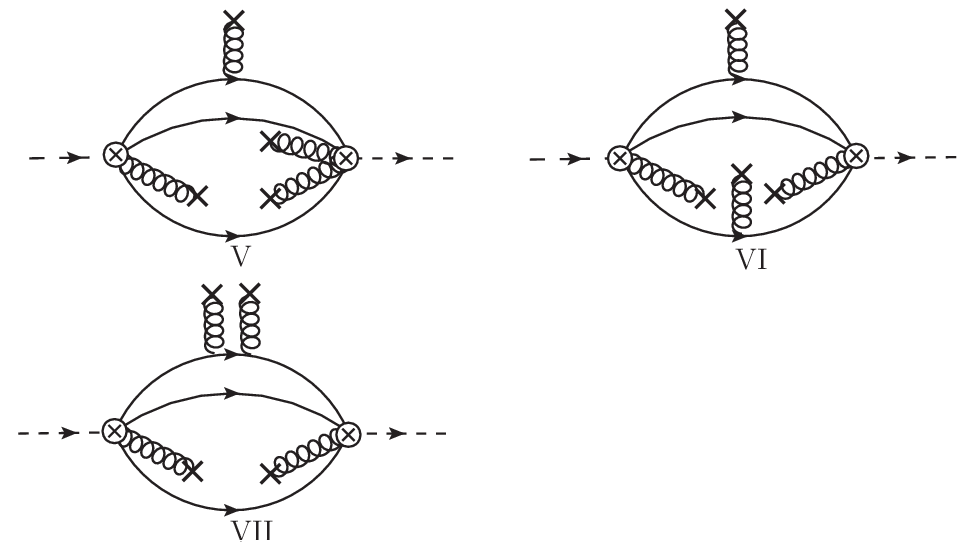}
\caption{The typical LO Feynman diagrams of a triply heavy hybrid baryon state contribute to the spectral densities in Eq.(\ref{spectral-density}), where the permutation diagrams are implied. Diagram I represents the contribution from the perturbative term, and diagrams II, III-IV, and V-VII denote the two-gluon condensate, three-gluon condensates, and four-gluon condensates, respectively.} \label{Feynman-diagram}
\end{center}
\end{figure}

The typical LO Feynman diagrams of a triply heavy hybrid baryon state that contribute to the spectral densities in Eq.(\ref{spectral-density}) are shown in Fig.~\ref{Feynman-diagram}, where diagram I denotes the contribution from the perturbative term, diagram II represents the two-gluon condensate, III and IV are the three-gluon condensates, and V-VII depict the four-gluon condensates, respectively.

After making the Borel transformation to Eq.(\ref{Pi-OPE}), we can reach the following equation
\begin{eqnarray}
  \Pi_{1}^{\text{OPE}}(M_B^2) = \int_{(3 m_Q)^2}^\infty ds \rho_{1}^{\text{OPE}}(s) e^{-s/M_B^2} . \label{Pi-MB}
\end{eqnarray}

For the triply heavy hybrid baryon states considered in this work, we put the lengthy expressions of spectral densities $\rho_{1}^{\text{OPE}}(s)$ appeared in Eq.(\ref{Pi-MB}) into the Appendix.

On the phenomenological side, after separating the ground state contribution from the pole term, the correlation function $\Pi(q)$ can be expressed as the dispersion integral over the physical region, \textit{i.e.},
\begin{eqnarray}
  \Pi(q) = \lambda_{H}\frac{q\!\!\!\slash + M_H}{(M_{H})^{2} - q^2} + \frac{1}{\pi} \int_{s_0}^\infty ds \frac{\rho^{\text{phen}}(s)}{s - q^2} + \text{subtractions}, \label{Pi-hadron}
\end{eqnarray}
where the subscript $H$ means the lowest-lying triply heavy hybrid baryon state, $M_{H}$ denotes its mass, and $\rho^{\text{phen}}(s) = q\!\!\!\slash \rho_1^{\text{phen}}(s) + \rho_2^{\text{phen}}(s)$ is the spectral density that contains the contribution from higher excited states and the continuum states above the threshold $s_0$. The coupling constant $\lambda_{H}$ is defined through
\begin{eqnarray}
\langle 0|j_{\Omega_H}(x)|H\rangle &=& \lambda_{H},
\end{eqnarray}
where $|H\rangle$ stands for the hybrid baryon state. It should be noted that, to obtain the expression shown in Eq.(\ref{Pi-hadron}), we have used the Dirac spinor sum relation,
\begin{eqnarray}
  \sum_s N(q,s) \bar{N}(q,s) = q\!\!\!\slash + M_H,
\end{eqnarray}
for spin-$\frac{1}{2}$ baryon.

By performing the Borel transformation on the phenomenological side, Eq.(\ref{Pi-hadron}), we have
\begin{eqnarray}
  \Pi^{\text{phen}}(M_B^2)=\lambda_H^2 (q\!\!\!\slash + M_H) e^{-M_H^2/M_B^2}+ \frac{1}{\pi} \int_{s_0}^\infty ds \rho^{\text{phen}}(s) e^{-s/M_B^2} .
\end{eqnarray}

According to the quark-hadron duality, the correlation functions obtained at the hadronic and quark-gluonic levels must equal to each other, based on which one can establish QCD sum rules relating these two levles. Thus, the QCD sum rules corresponding to $\Pi_1(q^2)$ reads as a function of the continuum threshold $s_0$ and Borel parameter $M_B^2$,
\begin{eqnarray}
  \lambda_H^2 e^{-M_H^2/M_B^2} &\!\!=\!\!& \int_{(3m_Q)^2}^{s_0} ds \rho_1^{\text{OPE}}(s) e^{-s/M_B^2}.
\end{eqnarray}

Eventually, we can extract the mass of the triply heavy hybrid baryon from the ratio of the derivative of the sum rules on $\frac{1}{M_B^2}$ and itself, and yields
\begin{eqnarray}
  M_H = \sqrt{-\frac{L_1^{\text{deri}}(s_0, M_B^2)}{L_1(s_0, M_B^2)}}\, ,
\end{eqnarray}
where the momentum $L_1(s_0, M_B^2)$ and its derivative $L_1^{\text{deri}}(s_0, M_B^2)$ are respectively defined as
\begin{eqnarray}
  L_1(s_0, M_B^2) &\!\!=\!\!& \int_{(3 m_Q)^2}^{s_0} ds \, \rho^{\text{OPE}}_1(s) e^{-s/M_B^2}, \label{L0} \\
  L_1^{\text{deri}}(s_0, M_B^2) &\!\!=\!\!& \frac{\partial}{\partial (M_B^2)^{-1}} L_1(s_0, M_B^2).
\end{eqnarray}

One can also obtain similar sum rules corresponding to $\Pi_2(q^2)$, but we will not discuss it in this paper.

\section{Numerical Evaluation}\label{Numerical}

In order to yield meaningful physical results in QCD sum rules, as in any practical theory, one needs to give certain inputs, such as the magnitudes of condensates and quark masses. These input parameters are taken from~\cite{ParticleDataGroup:2022pth, Narison:2011xe, Narison:2018dcr, Chen:2021smz, Tang:2021zti, Su:2022fqr}, whose explicit values read: $m_c (m_c) = \overline{m}_c= (1.27 \pm 0.02) \; \text{GeV}$, $m_b (m_b) = \overline{m}_b= (4.18^{+0.03}_{-0.02}) \; \text{GeV}$, where we use the ``running masses'' for the heavy quarks in the $\overline{\text{MS}}$ scheme; and for the two-gluon and three-gluon condensates, we take the prevailing values: $\langle \alpha_s G^2 \rangle = (6.35 \pm 0.35)\times  10^{-2} \; \text{GeV}^{4},\;\;\langle g_s^3 G^3 \rangle = (8.2\pm1.0)\times\langle \alpha_s G^2 \rangle \;$ $\text{GeV}^{2}$. It is important to note that the vacuum saturation approximation is used in this work in the calculation of $\langle G^4\rangle$ contribution~\cite{Shifman:1980ui, Bagan:1984zt}. In order to take into account the error due to the violation of the vacuum approximation, we can introduce a parameter $\kappa$,
\begin{eqnarray}
  \langle \alpha_s G^2 \rangle^2 \to \kappa \langle \alpha_s G^2 \rangle^2,
\end{eqnarray}
the value $\kappa =1$ stands for the vacuum saturation approximation, while the value $\kappa \neq 1$ parameterizes its violation. We consider the result obtained by using the factorized $\langle G^4 \rangle$ as the central value ($\kappa =1$) and consider the variation due to the violation of the vacuum dominance (by a factor of $\kappa =2$) as a source of errors.

For numerical evaluation, the leading order strong coupling constant
\begin{eqnarray}
\alpha_{s}(M_{B}^{2}) = \frac{4\pi}{(11-\frac{2}{3}n_{f})\ln(\frac{M_{B}^{2}}{\Lambda_{\text{QCD}}^{2}})},
\end{eqnarray} is adopted with $\Lambda_{\text{QCD}} = 300$ MeV and $n_{f} $ being the number of active quarks.

Furthermore, there exist two additional parameters $M_{B}^{2}$ and $s_{0}$ introduced in establishing the sum rule, which will be fixed in light of the so-called standard procedures abiding by two criteria~\cite{Shifman:1978bx, Shifman:1978by, Reinders:1984sr, P.Col, Narison:1989aq}. First, in order to extract the information on the ground state of the hybrid state, one should guarantee pole contribution (PC) is larger than 40\%, which can be formulated as
\begin{eqnarray}
  R_{1}^{\text{PC}} = \frac{L_1(s_0, M_B^2)}{L_1(\infty, M_B^2)} \; , \label{RatioPC}
\end{eqnarray}
where subscription 1 corresponds to the sum rules from $\Pi_1(q^2)$. Under this constraint, the contribution of higher excited and continuum states will be suppressed. This criterion gives rise to a critical value of $M_{B}^{2}$, which is the upper limit of $M_{B}^{2}$.

The second one asks for the convergence of the OPE. That is, one needs to compare individual contributions with the total magnitude on the OPE side and choose a reliable region for $M_{B}^{2}$ to retain the convergence. In practice, the degree of convergence may be expressed in fractions of condensates over the total as
\begin{eqnarray}
  R_{1}^{\text{cond}} = \frac{L_1^{\text{cond}}(s_0, M_B^2)}{L_1(s_0, M_B^2)}\,,
\end{eqnarray}
where the superscript `cond' denotes the perturbative term and different condensate terms in Eq.\eqref{L0}, respectively. Under this prerequisite, one can obtain the lower limit of $M_B^2$ called $(M_B^2)_{\text{min}}$, which ensures the contributions of higher dimension condensates are properly suppressed. As a consequence, we obtain the proper Borel window of $M_B^2$ for a given $s_0$, which is the region between $(M_B^2)_{min}$ and $(M_B^2)_{max}$.

In practice, to know whether the OPE convergence is satisfied, we firstly restrict that the highest condensate contribution, $\langle G^4 \rangle$, should be less than 5\% and 10\% of the total OPE side for $\kappa =1$ and $\kappa=2$, respectively. Then, we can select the one which has an OPE series decreasing order by order.

The standard procedure of the QCD sum rules is to mutually vary the threshold value $s_{0}$ to find its proper value that satisfies the two aforementioned criteria and find a smooth plateau for $M_H$ on the curve of the $M_H-M_B^2$. On the smooth plateau, the hybrid mass $M_H$ should be in principle independent of the Borel parameter$M_{B}^{2}$, or at least only shows weak dependence.

\begin{figure}[htb]
\begin{center}
\includegraphics[width=7.5cm]{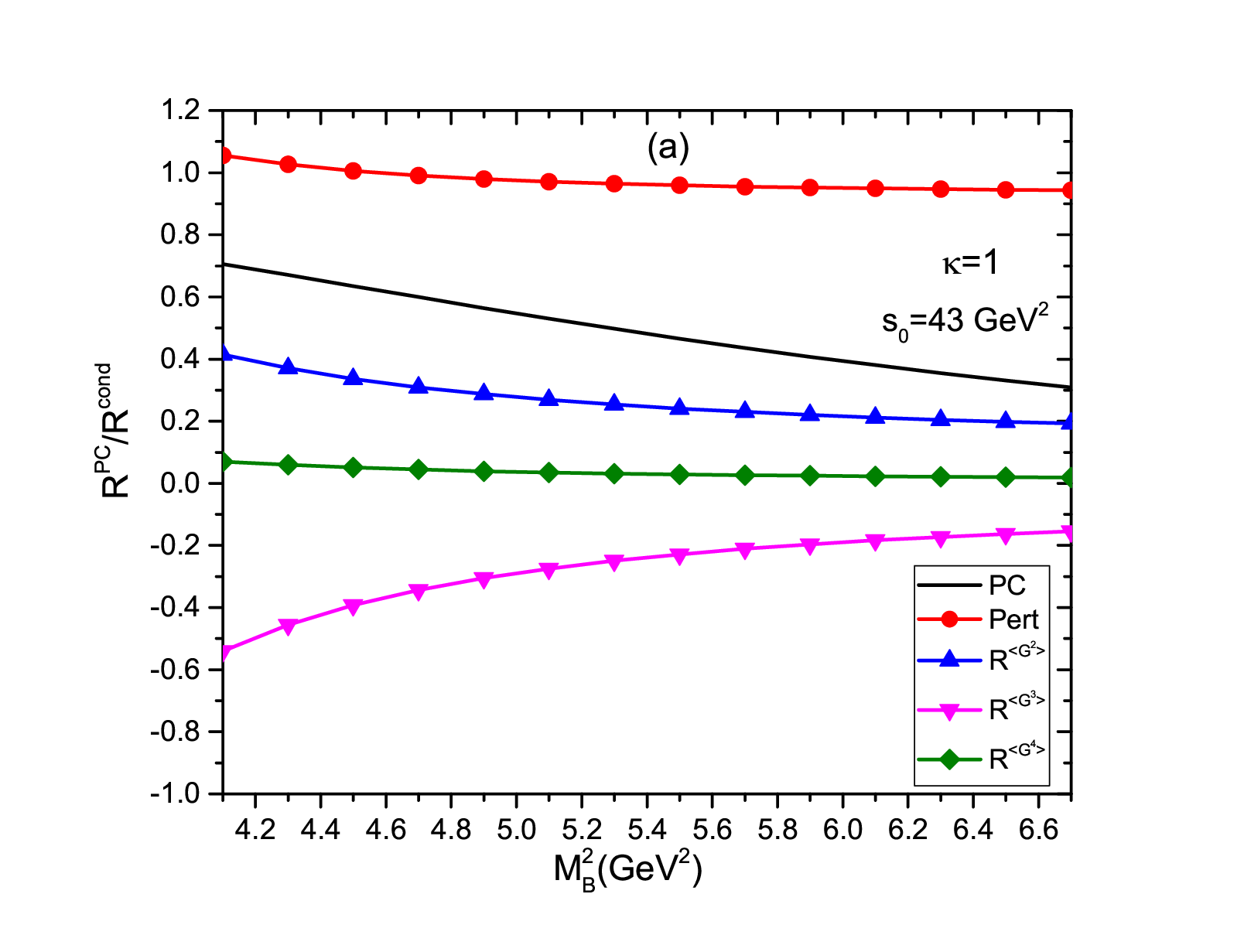}
\includegraphics[width=7.5cm]{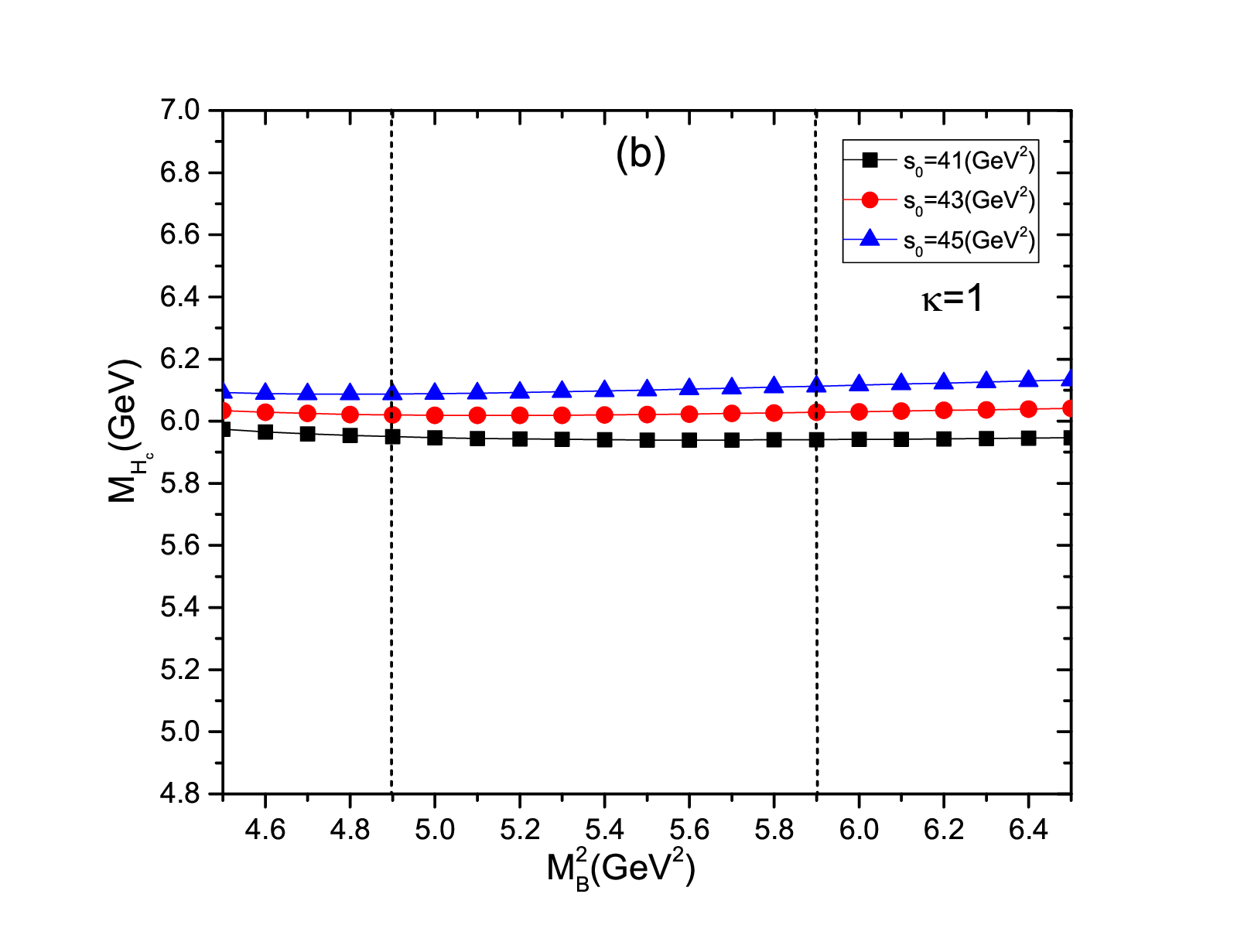}
\caption{The figures for the triply charmed hybrid baryon which is given in Eq.(\ref{current}) with $\kappa =1$. (a) The pole contribution ratio $R_{1}^{\text{PC}}$ and OPE convergence ratio $R_{1}^{\text{cond}}$ as functions of the Borel parameter $M^{2}_{B}$ with the central value of $s_{0}$; (b) The mass $M_{H}$ as a function of $M^{2}_{B}$ with different values of $s_{0}$, and the two vertical lines indicate the upper and lower bounds of valid Borel window with the central value of $s_{0}$.} \label{Ratio-mass-kappa-1}
\end{center}
\end{figure}

With the above preparation, we can numerically evaluate the mass spectrum of the triply charmed hybrid baryon. For $\kappa=1$, we plot the two ratios $R_{1}^{\text{PC}}$ and $R_{1}^{\text{cond}}$ as functions of the Borel parameter $M^{2}_{B}$ in Fig.~\ref{Ratio-mass-kappa-1}(a) at the proper values  $s_{0}= 43 \text{GeV}^{2}$, and the mass curves as functions of $M^{2}_{B}$ in Fig.~\ref{Ratio-mass-kappa-1}(b). Two vertical lines in Fig.~\ref{Ratio-mass-kappa-1}(b) indicate the upper and lower bounds of the valid Borel window for the central value of $s_{0}$, where the so-called stable plateau between these two vertical lines exists, suggesting the mass of possible hybrid baryon state. Here the valid Borel window refers to the one that not only requires the ratio $R_1^{\text{PC}}$ is larger than $40\%$ but also fulfills the constraints $|R_1^{\langle G^3 \rangle }| > |R_1^{\langle G^4 \rangle }|$ and $|R_1^{\langle G^3 \rangle} <30\%|$.

\begin{figure}[htb]
\begin{center}
\includegraphics[width=7.5cm]{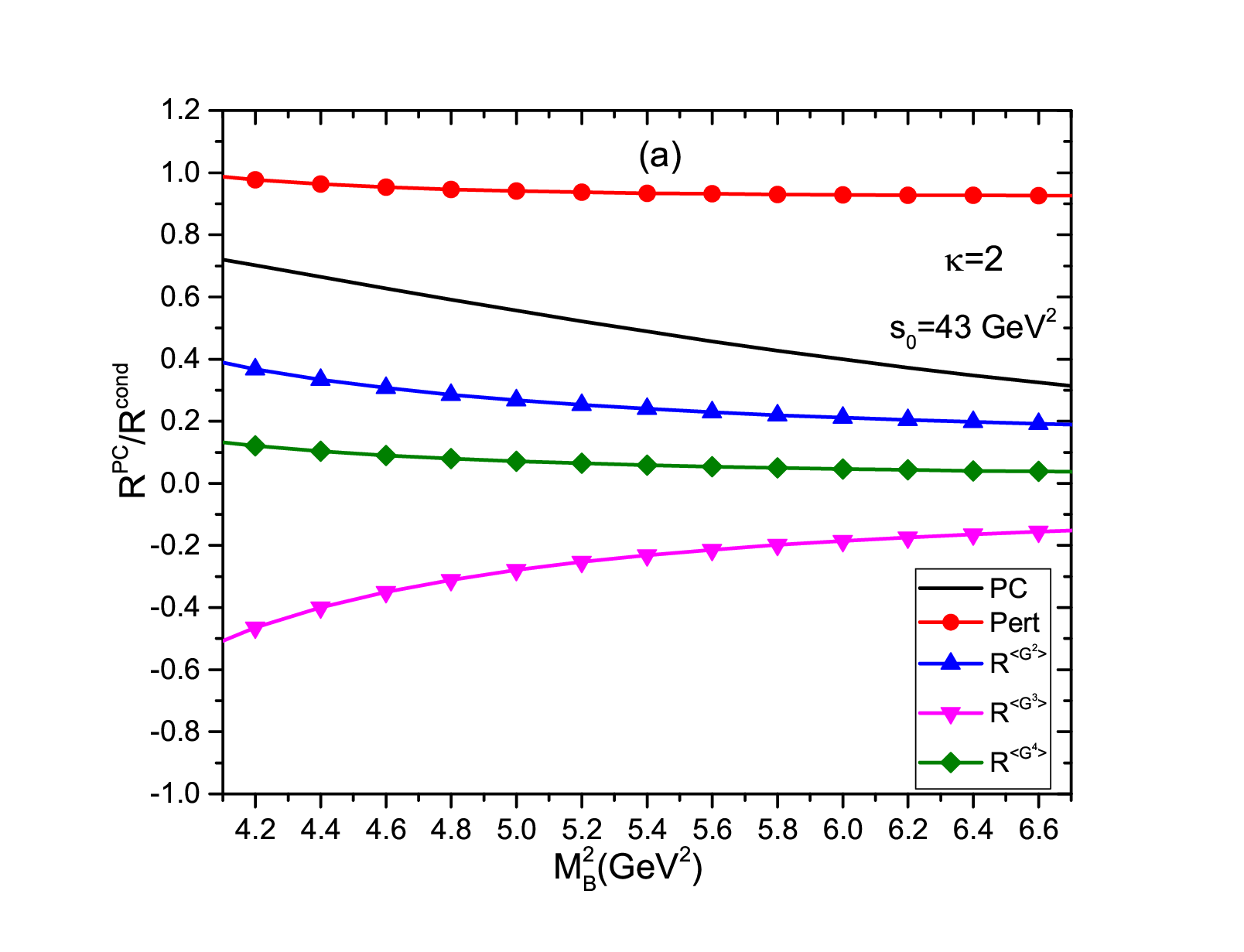}
\includegraphics[width=7.5cm]{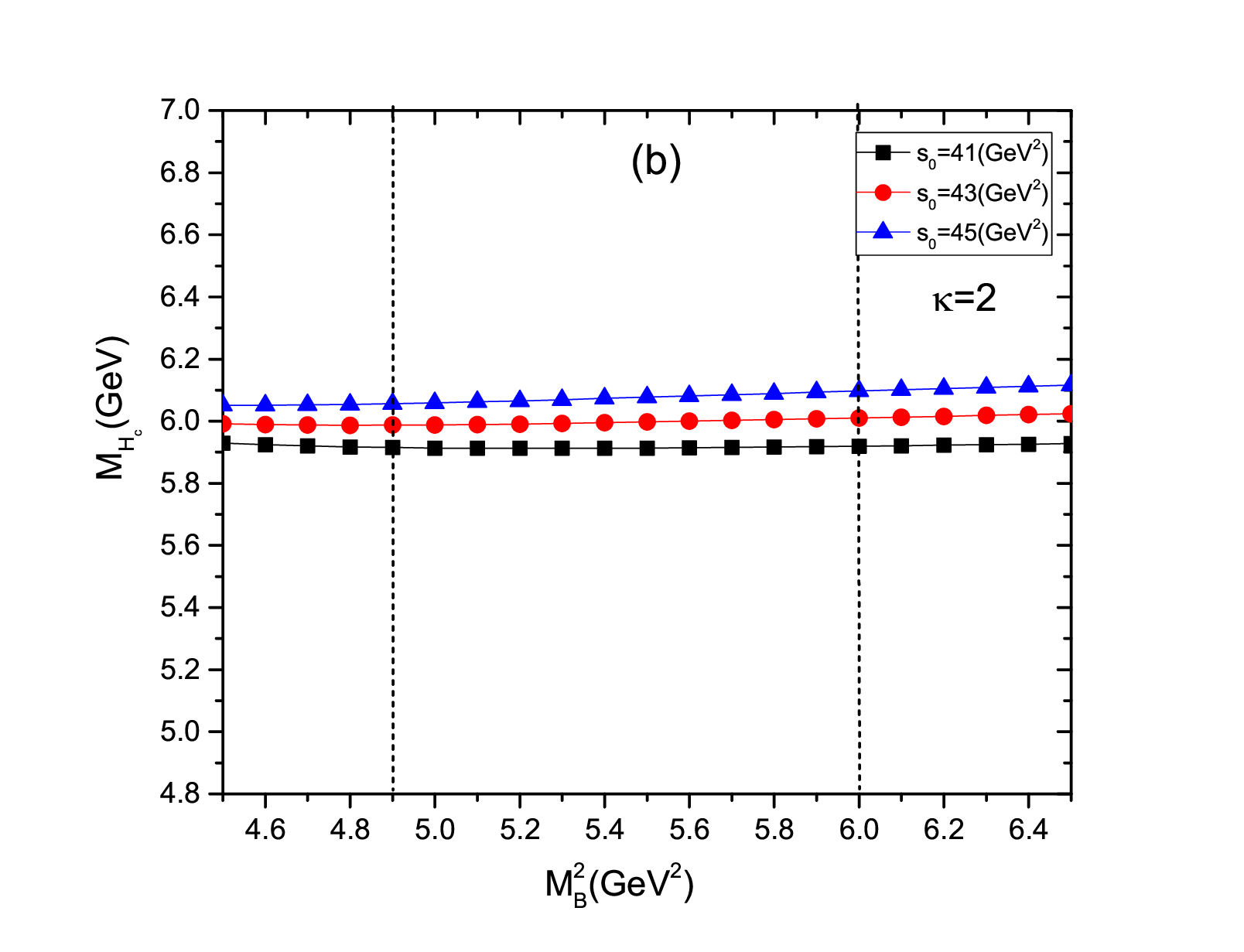}
\caption{The same caption as in Fig.~\ref{Ratio-mass-kappa-1}, but for $\kappa = 2$.} \label{Ratio-mass-kappa-2}
\end{center}
\end{figure}

A similar situation for $\kappa = 2$ is respectively shown in Fig.~\ref{Ratio-mass-kappa-2}, where a similar stable plateau between the Borel windows exists.

As already mentioned, we need to consider the variation due to the violation of the
vacuum dominance (by a factor of $\kappa = 2$) as a source of errors. Therefore, we summarize the corresponding results with $\kappa =2 $ in the second line of Table.~\ref{tab1}.
\begin{table}[htb]
\begin{center}
\begin{tabular}{|c|c|c|c|c|c|c|c|c}\hline\hline
  & $M_B^2 (\rm{GeV}^2)$  & $s_{0} (\rm{GeV^{2}})$ & PC & Pert &  $R_{1}^{\langle G^2 \rangle}$   &  $R_{1}^{\langle G^3 \rangle}$ &  $R_{1}^{\langle G^4 \rangle}$  \\ \hline
 $\kappa=1$ & $4.90\!-\!5.90$ & $43$ & $(56\!-\!40)\%$  & $(98\!-\!95)\%$ &$(29\!-\!22)\%$   & $[(-31)\!-\!(-20)]\%$  & $(4\!-\!3)\%$   \\
 \hline
 $\kappa=2$ & $4.86\!-\!6.00$ & 43 & $(58\!-\!40)\%$  & $(94\!-\!93)\%$ &$(28\!-\!21)\%$   & $[(-30)\!-\!(-19)]\%$  & $(8\!-\!5)\%$   \\
 \hline
 \hline
\end{tabular}
\end{center}
\caption{The windows of the Borel parameters $M^{2}_{B}$, threshold parameter $s_{0}$, pole contributions, perturbative contribution, two-gluon contributions, three-gluon contributions, and four-gluon contributions of the triply charmed hybrid baryon for $\kappa = 1$ and $\kappa =2 $, respectively.}
\label{tab1}
\end{table}

The resulting windows of the Borel parameters, threshold values $s_0$, pole contribution (PC), perturbative contribution, two-gluon contributions, three-gluon contributions, four-gluon contributions both for $\kappa =1 $ and $\kappa =2$ are shown explicitly in Table.~\ref{tab1}, respectively. From Table.~\ref{tab1}, we can see that both the pole dominance at the phenomenological side and the OPE convergence are well satisfied. In the Borel windows, the contributions of the two-gluon condensate, three-gluon contributions, four-gluon contributions are smaller than the total contribution. Moreover, the mass curves in these Borel windows have the optimal platform. Now the three requirements of the QCD sum rules are all satisfied, we expect to make reasonable predictions, which are summarizes as follows:
\begin{eqnarray}
  M_{H} &=& \left(6.02^{+0.11}_{-0.08}\right)\, \text{GeV} \;, \text{for} \; \kappa =1; \label{masskappa1} \\
  M_{H} &=& \left(6.00^{+0.11}_{-0.09}\right)\, \text{GeV} \;, \text{for} \; \kappa =2. \label{masskappa2}
\end{eqnarray}

After the above numerical analyses, we can then predict the masses of the low-lying triply charmed hybrid baryon with current Eq.(\ref{current}) for both $\kappa =1$ and $\kappa =2$, which are respectively shown in Eq.(\ref{masskappa1}) and Eq.(\ref{masskappa2}), where the central values correspond to the results with the optimal stability of $M^{2}_{B}$, and the errors stem from the uncertainties of the condensates, the quark mass, the threshold parameter $s_{0}$, and the Borel parameter $M^{2}_{B}$.

Eventually, by considering all the uncertainties mentioned above, we obtain the mass prediction of the triply charmed hybrid baryon state, which is
\begin{eqnarray}
M_{H} &=& \left(6.02^{+0.11}_{-0.11}\right) \, \text{GeV} \; ,
 \label{eq-mass-1}
\end{eqnarray}
and find that it is in the region of $5.91 \, \text{GeV} < M_{H}<6.13 \, \text{GeV}$.
By replacing the mass of the charm quark with the bottom quark in Eq.(\ref{Pi-OPE}) and performing the same numerical analysis, we can obtain the corresponding prediction for the triply bottom hybrid baryon state, whose mass is
\begin{eqnarray}
M_{H_b} &=& \left(14.68^{+0.14}_{-0.06}\right) \, \text{GeV} \; ,
 \label{eq-mass-2}
\end{eqnarray}
where the subscript $H_{b}$ represents the hybrid baryon state in b-quark sector. By including the uncertainties mentioned above, we conclude that it is in the range of $14.62 \, \text{GeV}< M_{H_{b}} < 14.82 \, \text{GeV}$.

\section{Decay analyses}\label{decay}

\begin{figure}[htb]
\begin{center}
\includegraphics[width=13cm]{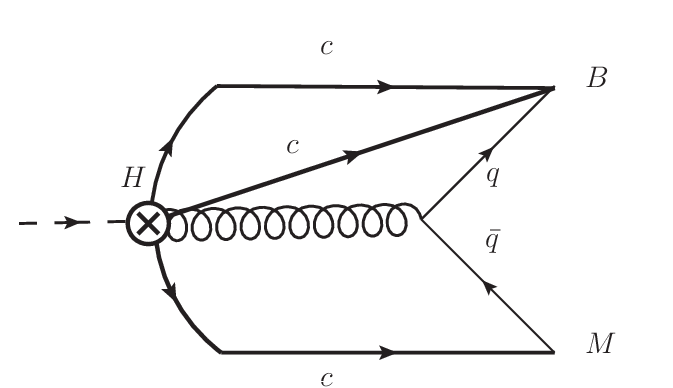}
\caption{A possible decay process of the triply charmed hybrid baryon $H$, where $q$ denotes the light quarks and the final states $B$ and $M$ represent the doubly charmed baryon and the charmed meson, respectively. }
\label{hybrid-baryon-decay}
\end{center}
\end{figure}

As shown in Fig.~\ref{hybrid-baryon-decay}, the triply charmed hybrid baryon can decay into one doubly charmed baryon and one  charmed meson by exciting one light quark ($u$, $d$, or $s$) pair from the valence gluon. It should be noted that this possible decay mode is at $\cal{O}$$(\alpha_s)^{1/2}$ order, though it is an OZI-allowed process.

\begin{table}[hbt]
\caption{Some possible one baryon and one meson decay channels of the $\Omega_{cccg}$ hybrid baryon with the quantum number $J^{P} = (1/2)^{+}$, where we only keep the channels up to P-wave decays.~\cite{ParticleDataGroup:2022pth}}
\begin{tabular}{|c|c|c|}\hline
\hline
                   & S-wave   & P-wave   \\ \hline
\multirow{1}{*}{$q=u$}& $\Xi_{cc}^{++} D_0^*(2300)^0$, $\Xi_{cc}^{++} D_1(2420)^0$, , $\Xi_{cc}^{++} D_1(2430)^0$  & $\Xi_{cc}^{++}  D^0$, $\Xi_{cc}^{++}  D^*(2007)^0$  \\ \hline
\multirow{1}{*}{$q=d$}& $\Xi_{cc}^{+} D_0^*(2300)^+$, $\Xi_{cc}^{+} D_1(2420)^+$  & $\Xi_{cc}^{+}  D^+$, $\Xi_{cc}^{+}  D^*(2010)^+$  \\ \hline
\multirow{1}{*}{$q=s$}& $\Xi_{ccs}^{+}  D_{s0}^*(2317)^+$, $\Xi_{ccs}^{+}  D_{s1}^*(2460)^+$ & $\Xi_{ccs}^{+}  D_s^+$, $\Xi_{ccs}^{+}  D_s^{*+}$  \\ \hline
\hline
\end{tabular}\label{tableII}
\end{table}

As shown in Table.~\ref{tableII}, both S-wave and P-wave decay modes exist for the triply charmed hybrid baryon $\Omega_{cccg}$. Amongst them that are listed in Table.~\ref{tableII}, we suggest the P-wave decay channels $\Xi_{cc}^{++}  D^0$, $\Xi_{cc}^{+}  D^+$, and $\Xi_{ccs}^{+}  D_s^+$ as the accessible decay channels for the triply charmed hybrid baryon $\Omega_{cccg}$, because they can be more easily produced in experiments. These decay channels are expected to be measured in BelleII, PANDA, Super-B, and LHCb in the near future.

\section{Conclusions}
This is the first study on the triply heavy hybrid baryons. We use the QCD sum rule method to evaluate their masses. Firstly, we construct the simplest hybrid baryon current, which has been extensively used in the studies of the lowest-lying light hybrid states~\cite{Kisslinger:1995yw, Kisslinger:2003hk, Azizi:2017xyx}. Then, we use this current to perform QCD sum rule analyses. The contributions up to dimension eight at the leading order of $\alpha_s$ (LO) in the operator product expansion are taken into account in the calculation. Eventually, we find that the mass of $cccg$ hybrid baryon lies in $M_{cccg}= 5.91-6.13$ GeV. As a byproduct, the mass of the triply bottom hybrid baryon state is extracted to be around $M_{bbbg}=14.62-14.82$ GeV.

The triply charmed hybrid baryon predicted in this work can decay into one doubly charmed baryon and one charmed meson. We list some possible decay patterns of the triply charmed hybrid baryon with the quantum numbers $J^P=(1/2)^+$ in Table~\ref{tableII}, where both the S-wave and P-wave decay channels are taken into account. As the charmed mesons presented in the P-wave decay modes are more plentiful in experiments than that shown in the S-wave decay channels, we conclude that the P-wave decay channels are more worthy to be searched for. Especially, we propose searching for the $J^{P}= (1/2)^+$ $cccg$ hybrid baryon in its P-wave decay channels $\Xi_{cc}^{++}  D^0$, $\Xi_{cc}^{+}  D^+$, and $\Xi_{ccs}^{+}  D_s^+$ in the future BelleII, Super-
B, PANDA, and LHCb experiments.

The present theoretical formalism is also expected to be extended to the bottom-charmed and heavy-light sectors to predict their mass spectra. We will analyze these hybrid baryons in detail in our next work.

\vspace{.7cm} {\bf Acknowledgments} \vspace{.3cm}

This work was supported in part by the Science Foundation of Hebei Normal University.


\begin{widetext}

\newpage

\appendix

\textbf{Appendix}

We list in this appendix the explicit expressions of the QCD spectral densities $\rho_1^{\text{OPE}}(s)$ in Eq.(\ref{Pi-MB}) for the current shown in Eq.~(\ref{current}).

For the $QQQg$ hybrid baryon states with $Q=c, b$, the spectral densities $\rho_1^{\text{OPE}}(s)$ are
\begin{eqnarray}
\rho^{\text{pert}, \text{I}}(s) &=& \frac{g_{s}^{2}}{2^{8}\times\pi^{6}}  \int_{\alpha_{min}}^{\alpha_{max}} d\alpha
\int_{\beta_{min}}^{\beta_{max}} d\beta
\int_{\gamma_{min}}^{\gamma_{max}} d\gamma \left\{ F_{\alpha\beta\gamma}^{4}\alpha\beta\gamma \right.\nonumber\\
&+&\left.2m_{c}^{2}\gamma F_{\alpha\beta\gamma}^{3}(\alpha+
\beta+\gamma-1)\right\},
\end{eqnarray}

\begin{eqnarray}
\rho^{\langle G^{2}\rangle, \text{II}}(s) &=&\frac{m_{c}^{2}\langle g_{s}^{2}G^{2}\rangle}
{2^{7}\times\pi^{4}\times\alpha\beta} \int_{\alpha_{min}}^{\alpha_{max}} d\alpha\int_{\beta_{min}}^{\beta_{max}} d\beta\left\{ F_{ \alpha\beta}(\alpha+\beta-1)\right\},
\end{eqnarray}

\begin{eqnarray}
\rho^{\langle G^{3} \rangle , \text{III}}(s) &=&\frac{\langle g_{s}^{3}G^{3}\rangle}
{2^{8}\times\pi^{4}} \int_{\alpha_{min}}^{\alpha_{max}} d\alpha\int_{\beta_{min}}^{\beta_{max}} d\beta\left\{(\alpha+\beta-1)(3m_{c}^{2}+8s\alpha\beta)\right.\nonumber\\
&-&\left.12(\alpha+\beta-1)F_{\alpha\beta}\right\} + \rho_{\delta}^{\langle G^{3} \rangle , \text{III}}(s),
\end{eqnarray}

\begin{eqnarray}
\rho_{\delta}^{\langle G^{3} \rangle , \text{III}}(s) &=&\frac{m_{c}^{4}\langle g_{s}^{3}G^{3}\rangle}
{2^{8}\times\pi^{4}}\int_{0}^{1} d\alpha\int_{0}^{1-\alpha} d\beta \delta(s - c_{\alpha \beta}m_Q^2)\times\left\{
\frac{\alpha^{4}+\alpha^{3}(2\beta-1)}
{\alpha\beta(\alpha+\beta-1)}\right.\nonumber\\
&+&\left.\frac{\alpha^{2}(3\beta^{2}-2\beta-1)+
\alpha(2\beta^{3}-2\beta^{2}-\beta+1)+(\beta-1)^{2}\beta(\beta+1)}
{\alpha\beta(\alpha+\beta-1)}\right\},
\end{eqnarray}

\begin{eqnarray}
\rho^{\langle G^{3} \rangle , \text{IV}}(s) &=& \frac{ m_{c}^{2}\langle g_{s}^{3}G^{3}\rangle}
{2^{9}\times\pi^{4}\alpha\beta} \int_{\alpha_{min}}^{\alpha_{max}} d\alpha\int_{\beta_{min}}^{\beta_{max}} d\beta\left\{ \alpha-\alpha^{2}-(\beta-1)\beta\right\},
\end{eqnarray}

\begin{eqnarray}
\rho^{\langle G^{4} \rangle , \text{V}}(s) &=&\frac{\kappa m_{c}^{2}\langle g_{s}^{2}G^{2}\rangle^{2}}
{2^{14}\times\pi^{4}}\int_{0}^{1} d\alpha\int_{0}^{1-\alpha} d\beta \delta(s - c_{\alpha \beta} m_Q^2) \times(\alpha+\beta)\left\{
\frac{2\alpha\beta(\alpha+\beta-1)}
{\alpha^{2}\beta^{2}}\right.\nonumber\\
&+&\left.\frac{m_{c}^{2}(\alpha^{2}+\alpha(\beta-1)+\beta(\beta-1))}
{M_{B}^{2}\alpha^{2}\beta^{2}}\right\},
\end{eqnarray}

\begin{eqnarray}
\rho^{\langle G^{4} \rangle , \text{VI}}(s) &=&\frac{31\kappa\langle g_{s}^{2}G^{2}\rangle^{2}}
{3\times2^{14}\times\pi^{4}} \int_{\alpha_{min}}^{\alpha_{max}} d\alpha\int_{\beta_{min}}^{\beta_{max}} d\beta
\left\{(\alpha+\beta)\right\} + \rho_{\delta}^{\langle G^{4} \rangle , \text{VI}}(s),
\end{eqnarray}

\begin{eqnarray}
\rho_{\delta}^{\langle G^{4} \rangle , \text{VI}}(s) &=&\frac{\kappa m_{c}^{2}\langle g_{s}^{2}G^{2}\rangle^{2}}
{9\times2^{14}\times\pi^{4}}\int_{0}^{1} d\alpha\int_{0}^{1-\alpha} d\beta \delta(s - c_{\alpha \beta} m_Q^2) \times\left\{
\frac{(-24+31\alpha^{3}+\beta-8\beta^{2})}
{\alpha\beta(-1+\alpha+\beta)}\right.\nonumber\\
&+&\left.\frac{31\beta^{3}+\alpha^{2}(-8+62\beta)+\alpha(1-16\beta+62\beta^{2})}
{\alpha\beta(-1+\alpha+\beta)}\right\},
\end{eqnarray}

\begin{eqnarray}
\rho^{\langle G^{4} \rangle , \text{VII}}(s) &=& \frac{\kappa \langle g_{s}^{2}G^{2}\rangle^{2}}
{27\times2^{14}\times\pi^{4}} \int_{\alpha_{min}}^{\alpha_{max}} d\alpha\int_{\beta_{min}}^{\beta_{max}} d\beta
\left\{\frac{\alpha^{4}(64\beta+15)}
{\alpha\beta(1-\alpha-\beta)}\right.\nonumber\\
&+&\left.\frac{\alpha^{2}(-64\beta^{3}+252\beta^{2}+34\beta+15)-2\alpha^{3}
(32\beta^{2}+49\beta+15)}
{\alpha\beta(1-\alpha-\beta)}\right.\nonumber\\
&+&\left.\frac{2\alpha\beta^{2}(32\beta^{2}-49\beta+17)+15(\beta-1)^{2}\beta^{2}}
{\alpha\beta(1-\alpha-\beta)}\right\} + \rho_{\delta}^{\langle G^{4} \rangle , \text{VII}}(s),
\end{eqnarray}

\begin{eqnarray}
\rho_{\delta}^{\langle G^{4} \rangle , \text{VII}}(s) &=& \frac{\kappa m_{c}^{2}\langle g_{s}^{2}G^{2}\rangle^{2}}
{81\times2^{15}\times\pi^{4}}\int_{0}^{1} d\alpha\int_{0}^{1-\alpha} d\beta \delta(s - c_{\alpha \beta}) \times\left\{
\frac{2m_{c}^{2}\times8\alpha^{8}(1+2\beta)}
{M_{B}^{2}\alpha^{3}\beta^{3}(1-\alpha-\beta)^{3}}\right.\nonumber\\
&+&\left.\frac{2m_{c}^{2}\times(\alpha^{7}(16\beta^{2}+20\beta-201)+4\alpha^{6}(55\beta^{2}
-223\beta+181))}
{M_{B}^{2}\alpha^{3}\beta^{3}(1-\alpha-\beta)^{3}}\right.\nonumber\\
&-&\left.\frac{2m_{c}^{2}\times2\alpha^{5}(16\beta^{4}-254\beta^{3}+925\beta^{2}
-1198\beta+523)}
{M_{B}^{2}\alpha^{3}\beta^{3}(1-\alpha-\beta)^{3}}\right.\nonumber\\
&+&\left.\frac{2m_{c}^{2}\times\alpha^{4}(-32\beta^{5}+696\beta^{4}-2123\beta^{3}
+3080\beta^{2}-2276\beta+684)}
{M_{B}^{2}\alpha^{3}\beta^{3}(1-\alpha-\beta)^{3}}\right.\nonumber\\
&+&\left.\frac{2m_{c}^{2}\times\alpha^{3}(508\beta^{5}-2123\beta^{4}+2618\beta^{3}
-1570\beta^{2}+736\beta-169)}
{M_{B}^{2}\alpha^{3}\beta^{3}(1-\alpha-\beta)^{3}}\right.\nonumber\\
&+&\left.\frac{2m_{c}^{2}\times2\alpha^{2}(\beta-1)^{2}\beta^{2}(8\beta^{3}+126\beta^{2}
-681\beta+52)}
{M_{B}^{2}\alpha^{3}\beta^{3}(1-\alpha-\beta)^{3}}\right.\nonumber\\
&+&\left.\frac{2m_{c}^{2}\times4\alpha(\beta-1)^{3}\beta^{3}(4\beta^{2}+17\beta-184)}
{M_{B}^{2}\alpha^{3}\beta^{3}(1-\alpha-\beta)^{3}}\right.\nonumber\\
&+&\left.\frac{2m_{c}^{2}\times(\beta-1)^{4}\beta^{3}(8\beta-169)}
{M_{B}^{2}\alpha^{3}\beta^{3}(1-\alpha-\beta)^{3}}\right.\nonumber\\
&+&\left.\frac{2\alpha^{7}(96\beta+31)+3\alpha^{6}(64\beta^{2}-104\beta+265)}
{\alpha^{2}\beta^{2}(1-\alpha-\beta)^{3}}\right.\nonumber\\
&+&\left.\frac{\alpha^{5}(-192\beta^{3}+956\beta^{2}+3718\beta-3800)}
{\alpha^{2}\beta^{2}(1-\alpha-\beta)^{3}}\right.\nonumber\\
&+&\left.\frac{\alpha^{4}(-384\beta^{4}
+2754\beta^{3}+4183\beta^{2}-11202\beta+6010)}
{\alpha^{2}\beta^{2}(1-\alpha-\beta)^{3}}\right.\nonumber\\
&-&\left.\frac{2\alpha^{3}(96\beta^{5}-1377\beta^{4}-1930\beta^{3}+6925\beta^{2}
-5769\beta+2055)}
{\alpha^{2}\beta^{2}(1-\alpha-\beta)^{3}}\right.\nonumber\\
&+&\left.\frac{\alpha^{2}(\beta-1)^{2}(192\beta^{4}+1340\beta^{3}+6671\beta^{2}-1848\beta+1043)}
{\alpha^{2}\beta^{2}(1-\alpha-\beta)^{3}}\right.\nonumber\\
&+&\left.\frac{2\alpha(\beta-1)^{3}\beta^{2}(96\beta^{2}+132\beta+1967)}
{\alpha^{2}\beta^{2}(1-\alpha-\beta)^{3}}\right.\nonumber\\
&+&\left.\frac{(\beta-1)^{4}\beta^{2}(62\beta+1043)}
{\alpha^{2}\beta^{2}(1-\alpha-\beta)^{3}}\right\}.
\end{eqnarray}

Here, we also have the following definitions:
\begin{eqnarray}
  \alpha_{min}=\frac{-3+\hat{s}-\sqrt{9-10\hat{s}+\hat{s}^{2}}}{2\hat{s}},
\end{eqnarray}

\begin{eqnarray}
      \alpha_{max}=\frac{-3+\hat{s}+\sqrt{9-10\hat{s}+\hat{s}^{2}}}{2\hat{s}},
\end{eqnarray}

\begin{eqnarray}
      \beta_{min}=\frac{1}{2(1-\hat{s})}(1-\alpha-\hat{s}\alpha+\hat{s}\alpha^{2}+\sqrt{((-1+\alpha+\hat{s}\alpha
    +\hat{s}\alpha^{2})^{2}-4(1-\hat{s}\alpha)(\alpha^{2}-\alpha))}),
\end{eqnarray}

\begin{eqnarray}
      \beta_{max}=\frac{1}{2(1-\hat{s})}(1-\alpha-\hat{s}\alpha+\hat{s}\alpha^{2}-\sqrt{((-1+\alpha+\hat{s}\alpha
    +\hat{s}\alpha^{2})^{2}-4(1-\hat{s}\alpha)(\alpha^{2}-\alpha))}),
\end{eqnarray}

\begin{eqnarray}
      \gamma_{min}=\frac{\alpha\beta}{-\alpha-\beta+\hat{s}\alpha\beta},
\end{eqnarray}

\begin{eqnarray}
      \gamma_{max}=1-\alpha-\beta.
\end{eqnarray}

It should be noted that for $QQQg$ case with $Q =c$ or $b$, we have the following definitions: $\hat{s} = \frac{s}{m_Q^2}$, ${\cal F}_{\alpha\beta\gamma} = m_Q^2 (\frac{1}{\alpha}+\frac{1}{\beta}+\frac{1}{\gamma}) - s$, ${\cal F}_{\alpha\beta} = m_Q^2(\beta+\frac{(-1+\alpha)\alpha}{-1+\alpha+\beta})-\alpha\beta s$, and $c_{\alpha\beta}=
\frac{1}{\alpha}+\frac{-1+\alpha}{\beta(-1+\alpha+\beta)}$.

\end{widetext}

\end{document}